\documentstyle[epsfig,mnras_cite,times]{mn}
\title{Fast Parameter Estimation from the CMB Power Spectrum}
\author[Sujata Gupta and Alan F. Heavens]
{Sujata Gupta and Alan F. Heavens\\
Institute for Astronomy, University of Edinburgh, Blackford Hill,
Edinburgh EH9 3HJ, U.K.}

\newcommand{\be}{\begin{equation}}
\newcommand{\ee}{\end{equation}}

\newcommand{\ba}{\begin{eqnarray}}
\newcommand{\ea}{\end{eqnarray}}

\newcommand{\lab}[1]{\label{eq:#1}}
\newcommand{\req}[1]{\ref{eq:#1}}

\newcommand{\bb}{{\bf b}}

\newcommand{\bx}{{\bf x}}

\def\gs{\mathrel{\raise1.16pt\hbox{$>$}\kern-7.0pt 
\lower3.06pt\hbox{{$\scriptstyle \sim$}}}}         
\def\ls{\mathrel{\raise1.16pt\hbox{$<$}\kern-7.0pt 
\lower3.06pt\hbox{{$\scriptstyle \sim$}}}}         

\begin{document}

\maketitle

\begin{abstract}
The statistical properties of a map of the primary fluctuations
in the cosmic microwave background (CMB) may be specified to high
accuracy by a few thousand power spectra measurements, provided
the fluctuations are gaussian, yet the number of parameters
relevant for the CMB is probably no more than about $10-20$.
There is consequently a large degree of redundancy in the power
spectrum data.  In this paper, we show that the MOPED data
compression technique can reduce the CMB power spectrum
measurements to about 10-20 numbers (one for each parameter),
from which the cosmological parameters can be estimated virtually
as accurately as from the complete power spectrum.  Combined with
recent advances in the speed of generation of theoretical power
spectra, this offers opportunities for very fast parameter
estimation from real and simulated CMB skies. The evaluation of
the likelihood itself, at Planck resolution, is speeded up by
factors up to $\sim 10^8$, ensuring that this step will not be the
dominant part of the data analysis pipeline.
\end{abstract}

\begin{keywords}
cosmic background radiation - cosmology; theory - early Universe
\end{keywords}

\section{Introduction}

It has been recognised for roughly a decade that detailed study
of the power spectrum of the fluctuations in the CMB could be
used to obtain high precision values for several of the
cosmological parameters, such as $\Omega_0$, $H_{0}$ and
$\Omega_\Lambda$ (\pcite{BE87}, \pcite{Kam94},
\pcite{Jungman96}). The physics of the CMB is much more
straightforward than the complicated processes which affect the
large-scale structure of the Universe, making it a much more
promising laboratory for accurate parameter estimation.  The main
complications are the presence of foreground sources at microwave
frequencies and proper accounting of instrumental noise effects,
but recent balloon experiments, Boomerang \cite{Boomerang}, MAXIMA
\cite{Maxima1} and DASI \cite{DASI1} have demonstrated that the
main scientific goal is achievable with current technology.  As
experiments become more ambitious, the data processing
requirements become more demanding, and the current datasets have
sufficiently many pixels ($\sim 10^4-10^5$) that the data
processing is already quite challenging. Even the first
measurement of the CMB fluctuations, produced by the Cosmic
Background Explorer (COBE) satellite (\pcite{COBE}) produced a
dataset with enough pixels ($\sim 4000$) for data compression
techniques to be valuable
(\pcite{Gorski94,Gorskietal94,Bond95,BunnSug95}).  For the
satellite experiments MAP (the Microwave Anisotropy Probe) and
Planck (the Planck Surveyor Satellite), data compression will be
vital. Each will provide very large datasets, with close to
all-sky coverage with a resolution of up to 5 arcminutes, and
$\sim 10^6 - 10^7$ pixels. The standard radical compression
method is to reduce the map to a set of power spectrum estimates
(see e.g. \pcite{BJK98}).  In principle this compression can be
lossless, if the map is a gaussian random field (as closely
predicted by inflation: see e.g. \pcite{GLMM94,VWHK00,WK00}), as
all the statistical properties of the map are calculable from the
power spectrum. The power spectrum data, typically a few thousand
numbers for a high-resolution experiment, can then be used to
estimate cosmological parameters to an accuracy of a few
percent.  The steps in the distillation of the raw data to the
cosmological parameters are, however, not necessarily
straightforward computationally (see e.g.
\pcite{Wright96,Muciaccia97,Teg97a,Teg97b,BCJK99,OSH99,Borrill99,Wandelt2000,Szapudi01,Natoli01,Hivon2001,Christensen2001}).
This paper addresses one aspect of this problem: parameter
estimation from the power spectrum.  MOPED\footnote{MOPED
(Multiple Optimised Parameter Estimation and Data Compression) has
patent protection} is an eigenvector-based method for data
compression and parameter estimation, originally developed for
computing star-formation histories from galaxy spectra (see
\pcite{HJL00}, hereafter HJL; \pcite{RJH01}). It can also be
employed for very accurate, and extremely fast, parameter
estimation from the CMB. The speed-up over brute-force maximum
likelihood method is dependent on the experiment: typical
speed-up factors expected for MAP and Planck are between $10^7$
and $10^9$. MOPED is much more powerful than necessary, in fact,
as parameter estimation will be dominated by the time it takes to
run predictions for cosmological models, or other steps in the
analysis pipeline.

The method is based on a technique developed by HJL for
compressing and analysing galaxy spectra.  In that paper, it was
shown that datasets with certain noise properties offered
possibilities for very radical linear compression of the data
without any loss of information about the parameters which
determine the data.  The requirement is for a dataset whose mean
depends on the parameters, but the covariance of the noise does
not.  In these circumstances, it is possible to find a set of
linear combinations of the data which are {\em locally sufficient
statistics} for the parameters - i.e. the compressed data contain
as much information about the parameters as the full dataset, and
in this sense the compression is lossless (strictly, the Fisher
matrix is unchanged, so the likelihood surface is known to be
unchanged only locally near the peak). The compressed dataset can
be extremely small - it consists of a single number for each
parameter. Thus for highly redundant datasets, the degree of
compression can be very large.

It is important to recognise that the data compression can still
be done even if the assumptions for lossless compression do not
apply. The main assumptions are that the information is contained
in the mean of the data, not in their variance, and that the
fiducial model is correct.  Violation of neither of these is
serious for CMB power spectrum analysis.  In HJL, for example,
the data compression algorithm was applied to the case of galaxy
spectra, where the noise includes a photon counting noise term
which is dependent on the mean number of photons in the spectral
channel, and hence does depend on the parameters of the galaxy.
The compressed data can still be used for parameter estimation,
but the error bars on the derived parameters are fractionally
larger than by using the full spectrum.  The same situation
arises in the CMB: under general assumptions, the cosmic variance
on a power measurement is proportional to the square of the power
itself, and therefore is dependent on the underlying parameters.
The data compression, although not lossless, is still highly
efficient: conditional errors should increase by a factor $\sim
1/\ell_{\rm max}$ for an experiment measuring multipoles up to
$\ell_{\rm max}$.  The time required for a brute-force likelihood
evaluation is broadly comparable to the time it takes to compute
theoretically the power spectrum of a model, using CMBFAST
\cite{SZ96}.  Significantly, this part of the process has been
accelerated recently by a factor of $\sim 10^3$ \cite{TZH01},
making it much faster to compute the theoretical power spectra
than computing a brute-force likelihood measurement..  The
relative timings for these two steps can determine the analysis
strategy, since if the computation of the theoretical power
spectrum is small in comparison with the likelihood evaluation,
on can calculate the power spectrum `on the fly' as one searches
through parameter space. A useful goal is therefore to make the
likelihood evaluation much quicker than computation of the
theoretical power spectrum. One can already speed up this process
by using variants of the Newton-Raphson method (see, e.g.
\pcite{BCJK99}), and one can argue that the power of MOPED is not
strictly necessary for this problem. However, it is possible that
calculations of theoretical power spectra will be accelerated
still further, but this paper shows that, with MOPED, the
analysis need never be dominated by likelihood evaluations.

In this paper, we demonstrate that MOPED does successfully
recover cosmological parameters from simulated datasets, but many
orders of magnitude more quickly. We also show that the parameter
errors are similar to the full maximum likelihood solution.

\section{Massive Lossless Data Compression}

The method is detailed in HJL, so we only sketch details here. We
define the data vector $\bx$ as the estimates of the power
spectrum $\{\hat{C_\ell}\}$, where $\ell$ is the angular
multipole, in terms of signal $C_\ell$ and noise $n_\ell$:
\begin{equation}
\hat{C_\ell} = C_\ell(\theta_\alpha) + n_\ell
\end{equation}
where {$\theta_\alpha$} are the set of cosmological parameters on
which the CMB power spectrum depends. The noise is assumed to
have zero mean, so
\begin{equation}
\langle\hat{C}_\ell\rangle=C_\ell(\theta_\alpha)
\end{equation}
and the noise covariance matrix, including cosmic variance and
instrument noise, is ${\mathcal N}_{\ell \ell'}=\, \langle n_{\ell}
n_{\ell'}\rangle$.  Angle brackets indicate ensemble averages; these
are calculable analytically for some algorithms of power spectrum
estimation (see e.g. \pcite{Teg97b}), but for others, e.g. based on
correlation functions \cite{Szapudi01}, a Monte Carlo approach is
required.  In practice this should be the covariance of the {\em
estimates} of the power spectrum.  Since this is dependent on the
algorithm used to estimate the power spectrum, we assume for
illustration only cosmic variance, modelled as gaussians with variance
$\frac{2C_\ell^2}{(2\ell+1)}$, but in addition we do correlate the
power spectrum estimates to mimic partial sky coverage.  This
approximation may not be good, especially for low multipoles.
\scite{BJK00} have argued that the distribution may be closer to an
offset lognormal, in which case one can transform the power spectrum
estimates to quantities which have nearly gaussian marginal
distributions.  The calculation we show is illustrative, but Planck
will be cosmic variance limited up to high $\ell$.

The brute force maximum likelihood method, which uses all the
power spectrum data points, is the method of estimation which for
a large dataset will provide the smallest errors, assuming
uniform priors. The likelihood for the {\small N} parameters is
\begin{eqnarray} \lab{full_lk}
{\mathcal L}(\theta_\alpha) &=&
\frac{1}{(2\pi)^\frac{\mbox{\scriptsize N}}{2}\sqrt{|{\mathcal
N}|}} \nonumber\\
& & \!\!\!\!\!\!\!\!\!\!\!\!\!\!\!\!\!\!\!
\!\!\!\!\!\!\!\!\!\!\!\!\!\!\!\!\exp\left\{-\frac{1}{2}\sum_{\ell
\ell'}
\left[\hat{C_\ell}-C_\ell(\theta_\alpha)\right]\,\,{\mathcal
N}_{\ell \ell'}^{-1}(\theta_\alpha)\,\,
\left[\hat{C_{\ell'}}-C_{\ell'}(\theta_\alpha)\right]\right\}
\end{eqnarray}

The difficulty is that at each point in parameter space one
generally computes the determinant of, and inverts, an ${\small
N} \times {\small N}$ matrix. Since this scales as {\small N}$^3$,
it becomes a significant computational expense, even with {\small
N} $\simeq 2000$.  In this context, significant means that it
exceeds significantly the time taken currently to generate the
theoretical power spectrum.

\begin{figure}
\centerline{\psfig{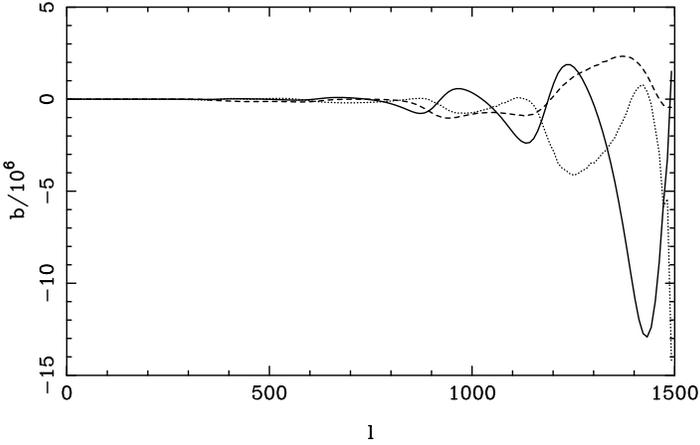}}
\caption{Optimised MOPED weighting vectors for a fiducial model
with $H_0=65$ km s$^{-1}$Mpc$^{-1}$, $\Omega_{\scriptsize
CDM}=0.254$ and $\Omega_\Lambda=0.7$.  The parameter ordering
(see text) is $\Omega_\Lambda$, $H_0$ and $\Omega_{\scriptsize
CDM}$, and the MOPED vectors ${\bf b}_1$, ${\bf b}_2$, ${\bf
b}_3$, are shown by the solid, dashed and dotted lines
respectively.  Derivatives of the power spectrum have been
calculated using finite-differences, which can cause the small
glitches seen in this figure.} \label{Evectors}
\end{figure}

We can speed up the likelihood evaluation by using MOPED to compress
the {\small N} data in the measured $\hat{C_\ell}$ to one datum for
each of $M$ unknown parameters. The algorithm is detailed in HJL; it
produces a set of weighting vectors $\bb_\alpha\ (\alpha=1\ldots M)$,
from which a set of MOPED components
\begin{equation}
y_\alpha \equiv
\bb_\alpha\cdot\bx
\end{equation}
is constructed.  The MOPED vectors are designed to
make the Fisher information matrix \be F_{\alpha\beta} \equiv
-\left\langle{\partial^2\ln {\cal L}\over \partial \theta_\alpha
\partial \theta_\beta}\right\rangle \ee the same whether we use the
compressed data $y_\alpha$ or the full set of power spectrum
estimates.  In fact this is only possible if we ignore the dependence
of cosmic variance on the parameters, but this restriction makes
virtually no difference for a CMB dataset.  The MOPED vectors satisfy
the following (HJL equation 14):
\begin{equation}
\bb_{\ell 1} = {{\mathcal N}^{-1}_{\ell \ell'}\frac{\partial
C\!_{\ell '}}{\partial \theta_{1}}\over {\frac{\partial C\!_{\ell
'}}{\partial \theta_1}{\mathcal N}^{-1}_{\ell ' \ell
''}\frac{\partial C\!_{\ell ''}}{\partial \theta_1}}}
\end{equation}
\begin{equation}
\bb_{\ell \alpha}=\frac{{\mathcal N}^{-1}_{\ell \ell
'}\frac{\partial C\!_{\ell '}}{\partial \theta_{\alpha}} -
 \sum^{\alpha - 1}_{\beta = 1}\left(\frac{\partial C\!_{\ell
'}}{\partial \theta_{\alpha}}\bb_{\beta \ell '}\right){\mbox
b}_{\beta \ell}}{\sqrt{\frac{\partial C\!_{\ell '}}{\partial
\theta_\alpha}{\mathcal N}^{-1}_{\ell ' \ell ''}\frac{\partial
C\!_{\ell ''}}{\partial \theta_\alpha} - \sum^{\alpha - 1}_{\beta
= 1}\left(\frac{\partial C\!_{\ell '}}{\partial
\theta_{\alpha}}\bb_{\beta \ell '}\right)^{2}}}
\end{equation}
and the summation convention is assumed. $\bb_{\ell\alpha}$ refers
to the $\ell$ component of the vector labelled by $\alpha$.
Obvious modifications are made if the data does not include all
$\ell$ values - the vector components refer to the list of modes
considered. Note that the MOPED vectors depend on the order in
which the parameters are listed: $y_1$ contains as much
information about parameter 1 as possible.  This vector also
constrains parameter 2 to some extent; $y_2$ adds as much
additional information as possible about parameter 2, etc. A set
of 3 MOPED vectors is illustrated in Fig. \ref{Evectors},
corresponding to vacuum energy density, Hubble constant and cold
dark matter (CDM) density.
These vectors would ensure, under certain assumptions, that the MOPED
components $y_\alpha$ are uncorrelated, and of unit variance; if this
is the case, the likelihood with these as the data is simply
\begin{equation} \lab{comp_lk}
{\mathcal L}(\theta_{\alpha})=
\frac{1}{(2\pi)^{3/2}}\exp\left[-\frac{1}{2}\sum_{i = 1}^3(y_i -
\langle y_i\rangle)^{2}\right]
\end{equation}
where the $\langle y_i\rangle$ are computed from the noise-free
(but smoothed) theoretical power spectra.  Importantly, they
ensure that the Fisher matrix for the compressed dataset
$\{y_\alpha\}$ is the same as for the entire set of power
spectrum estimates. The marginal error on a single parameter is
$[(F^{-1})_{\alpha\alpha}]^{\frac{1}{2}}$ and the error on the
parameter estimated using any method cannot be smaller than this
(see e.g. \pcite{Kendall,TTH97}).  Thus, by ensuring that the
Fisher matrices coincide, the compression method can be described
as locally lossless - the parameter errors, as estimated from the
local curvature of the likelihood surface at the peak, are on
average no larger for the compressed data than for the full set
of power spectrum estimates.

In detail, the assumptions required for locally lossless
compression do not hold for this analysis.  In order to calculate
the MOPED vectors, the data covariance matrix, and the
derivatives of the power spectrum with respect to the parameters,
need to be known. These are fixed by assuming a fiducial set of
parameters.  We show below that this fiducial set is not
important, but one can iterate the process if desired, at minimal
extra computational expense.  Our results show that iteration is
actually unnecessary. The second assumption is that the
covariance matrix of the data is not dependent on the model
parameters. This is not strictly true for the CMB power spectrum,
as the noise includes a cosmic variance term which is dependent
on the cosmology. However, this does not prevent us compressing
the data, and, in fact the Fisher matrix is dominated by the
sensitivity of the power spectrum itself to the parameters,
rather than the sensitivity of the noise.

A few remarks on speed are in order.  With $N$ power spectrum
estimates, brute-force likelihood calculations require $O(N^3)$
calculations.  With $M$ parameters, MOPED requires $O(M)$ operations
per likelihood evaluation.  In addition, there are $O(MN)$ operations
to compute the $\langle y_i \rangle$ quantities, but these can be done
in advance if a library of theoretical power spectra is built up prior
to analysis of the data.  This point is potentially important for
Planck; libraries of theoretical power spectra (and $\langle y_i
\rangle$) can be constructed in the years before launch; if so, the
parameter estimation step can be a very fast process, utilising
interpolation of the $\langle y_i \rangle$ if desired.

In addition to this, there is a one-off $O(MN^3)$ operation to
compute the MOPED vectors.  The number of likelihood evaluations
required to find the maximum is not easy to compute a priori, but
is likely to depend exponentially on $M$, so for a
large-dimensional parameter space, the overhead in computing the
vectors is negligible in comparison with time spent in searching
the space.
\begin{figure}
\centerline{\psfig{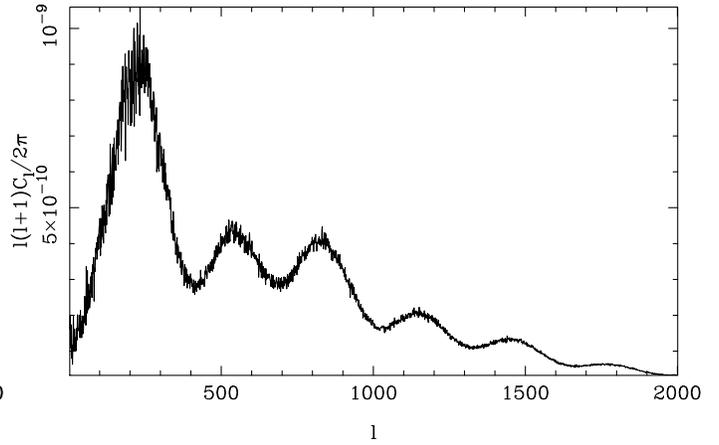}}
\caption{Simulated realisation of the CMB power used in the
analysis.}\label{noisy_spect}
\end{figure}

\section{Results}

\begin{figure}
\centerline{\psfig{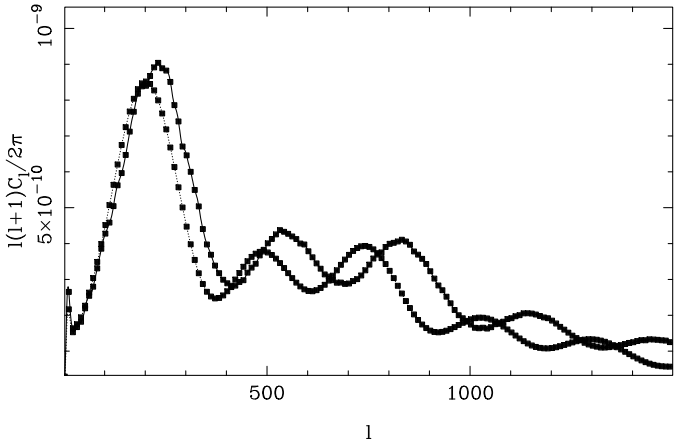}} \caption{The true model spectrum
(solid), with $H_0$ = 65 kms$^{-1}$Mpc$^{-1}$, $\Omega_\Lambda$ =
0.7 and $\Omega_{\mbox {\scriptsize CDM}}$ = 0.254, with gaussian
noise and smoothed in $\ell$ with a gaussian of width
$\Delta\ell=5$. Also shown (dotted) is the fiducial model used in
the data compression for fig.\ref{1500st10clk}: $H_0$ = 69.8
kms$^{-1}$Mpc$^{-1}$, $\Omega_\Lambda$ = 0.758 and $\Omega_{\mbox
{\scriptsize CDM}}$ = 0.254, both smoothed with a gaussian of
width $\Delta\ell=5$. The boxes show the data points used for the
likelihood calculations.}\label{truefid}
\end{figure}

We simulate a CMB dataset by adding gaussian noise, at the level
of cosmic variance, to theoretical power spectra produced by
CMBFAST. The power spectrum is convolved with a gaussian of
chosen width, to mimic approximately the correlations in power
spectrum estimates introduced by partial sky coverage. The
dataset consisted of the power spectrum sampled in even steps in
$\ell$.   The model chosen has $H_0=65$ kms$^{-1}$Mpc$^{-1}$,
$\Omega_\Lambda= 0.7$ and $\Omega_{\mbox {\scriptsize
CDM}}=0.254$. The unconvolved power spectrum is shown in fig.
\ref{noisy_spect}, and the convolved spectrum in fig.
\ref{truefid}.

We calculate the full (equation \req{full_lk}) and compressed
(equation \req{comp_lk}) likelihoods, varying the calculation in
the following ways:
\begin{itemize}
\item We mimic the effects of partial sky coverage by convolving the
power spectrum with a gaussian window function of various widths; we
present results for a width of $\Delta\ell = 5$.
\item The size of the dataset {\small N} is varied by changing the
upper multipole limit of the available data, or by missing out some
$C_\ell$ values.
\item We explore different fiducial models, to see if the method is
sensitive to an accurate initial guess of the parameters.
\end{itemize}
We fix most of the cosmological parameters.  The values are not
particularly important, but are listed here: $\Omega_B=0.05$;
scalar spectral index $n=1$; no tensor modes; no massive
neutrinos; 3 massless neutrinos.  The parameters we allow to vary
are the vacuum energy density parameter $\Omega_\Lambda$, the CDM
density parameter $\Omega_{\scriptsize CDM}$ and the Hubble
constant $H_0$, although we only display likelihood surfaces in
the $\Omega_\Lambda-H_0$ plane, with $\Omega_{\scriptsize CDM}$
fixed.

\begin{figure}
\centerline{\psfig{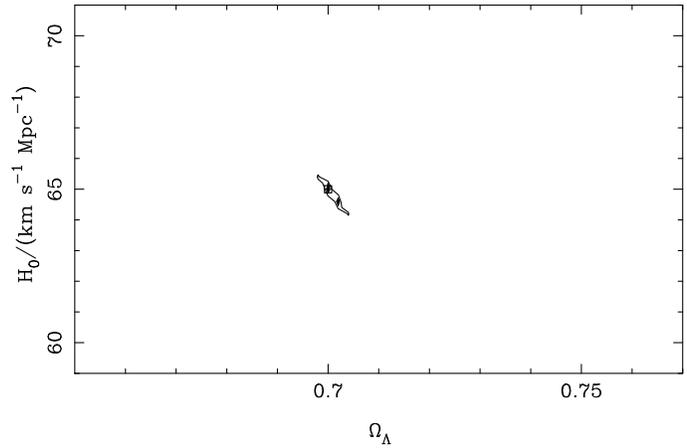}} \caption{Likelihood surface for
$\Omega_\Lambda$ and $H_0$ obtained from the the full dataset.
This dataset consists of 150 power spectrum estimates from $\ell
= 2,\ldots,1500$ in steps of 10, smoothed over a scale of
$\Delta\ell = 5$. The true model is labelled with a square
.  The likelihood contours are too small to see
individually for this experiment; the outer contour contains
99.99\% of the probability, assuming uniform priors.}\label{1500st10flk}
\end{figure}


\begin{figure}
\centerline{\psfig{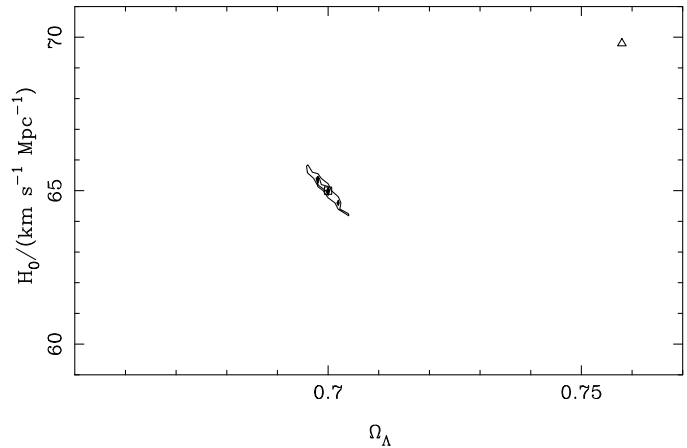}} \caption{Likelihood surface for
$\Omega_\Lambda$ and $H_0$ obtained from the the 3 MOPED
components.   The fiducial model used for the data compression no
longer coincides with the true model, and is marked by a
triangle.  Note that the method still recovers the correct model
(square). }\label{1500st10clk}
\end{figure}

\begin{figure}
\centerline{\psfig{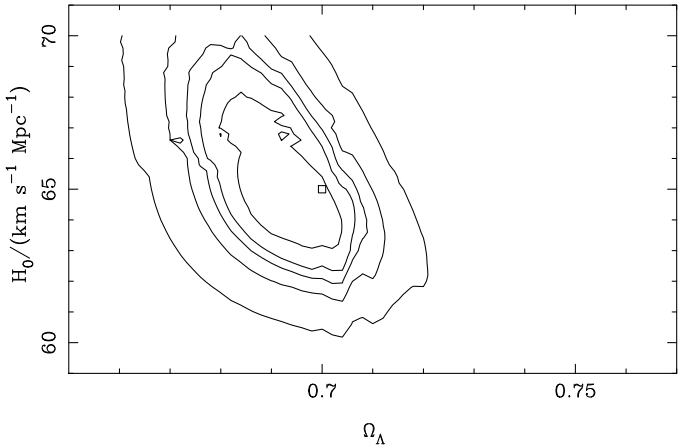}} \caption{Likelihood from the full
power spectrum, as in fig. \ref{1500st10flk}, but restricted to
$\ell\le 300$ in steps of 10, to illustrate the size of the error
bars.  The contours represent confidence limits of 99.99\%, 99\%,
95.4\%, 90\%,and 68\%.  The true model is labelled with a
square.  The likelihood was calculated on a grid
covering $60 \le H_0 \le 70$ km$\,s^{-1}\,Mpc^{-1}$ in steps of 0.2,
and $0.66 \le \Omega_\Lambda \le 0.76$ in steps of 0.002.}\label{300st10flk}
\end{figure}

\begin{figure}
\centerline{\psfig{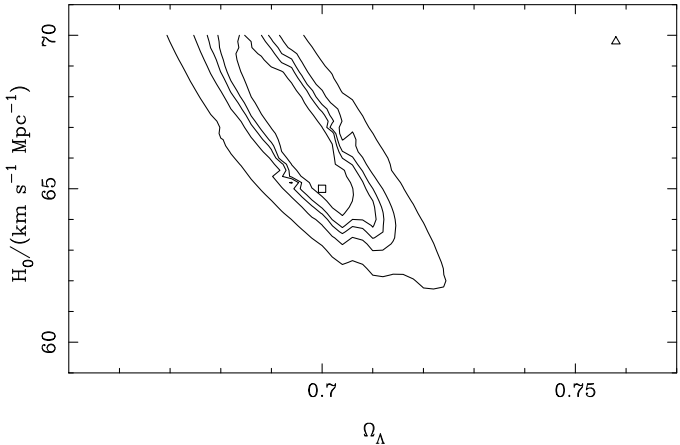}} \caption{As fig. \ref{300st10flk},
but showing the likelihood from MOPED components.  Note that the
error bars are comparable.}\label{300st10clk_fidmod}
\end{figure}

Figure \ref{1500st10flk} shows the $H_0-\Omega_\Lambda$ likelihood
surface using the power spectrum of Figure \ref{noisy_spect} up to
$\ell$ = 1500 in steps of 10. The power estimates were smoothed
with a gaussian of width 5. The calculation of this grid of
likelihoods took 9463 seconds of CPU on an alpha workstation.
Figure \ref{1500st10clk} shows the likelihood using 3 MOPED
components as compressed data.
An incorrect fiducial model ($H_0$ = 69.8 kms$^{-1}$Mpc$^{-1}$,
$\Omega_\Lambda$ = 0.758, $\Omega_{\mbox {\scriptsize CDM}}$ =
0.254) was chosen, to illustrate that its choice is not important.
The true solution is still recovered accurately,  but much
faster: 0.00098 seconds, or an improvement of order $10^7$.

In order to check that the compressed data recover the parameters
as accurately as the full data, we degrade the experiment,
truncating the data to $\ell=2,\ldots,300$, in steps of 10 (fig.
\ref{300st10flk} and \ref{300st10clk_fidmod}).   The method is
designed to ensure that the error bars should be almost the same
as the full likelihood on average, and we see that for this
realisation the errors are indeed comparable. The full likelihood
calculation takes 1406 seconds, while MOPED takes 0.00016
seconds.  We see here that with a very poor fiducial model, MOPED
still correctly finds the solution, within the errors, but there
is a suggestion that the errors are only approximately correct.
This can arise because the $y_i$ are assumed to be uncorrelated,
and this is only strictly true if the fiducial model is correct,
and even then it is only the ensemble average errors which are
unchanged. In practice, this is not a problem, as we have a much
better idea now of the shape of the power spectrum, so can choose
a fiducial model which is far better than this one. Secondly, one
can iterate, at very modest extra computational expense, computing
new MOPED vectors from the best previous estimate.

\section{Conclusions}

The steps required to turn a set of power spectrum measurements
$C_\ell$ into estimates of cosmological parameters consist of
\begin{itemize}
\item{Computation of theoretical $C_\ell$}
\item{Calculation of likelihood of model parameters}
\item{Maximisation of likelihood and marginalisation}
\end{itemize}
\scite{TZH01} have addressed the speed of the first step, accelerating
CMBFAST \cite{SZ96} by a factor $\sim10^3$.  This paper complements
that analysis by speeding up the brute-force likelihood evaluation in
the second step by even larger factors.  For $N$ correlated data
points, a brute-force likelihood evaluation using all the data scales
as $N^3$.  MOPED reduces this to $M$ approximately uncorrelated, unit
variance components, whose likelihood evaluation scales with the
number of parameters $M$.  For a Planck-size dataset with $N=2000$ and
$M \sim 12$ parameters, the speed-up factor should be around 500
million. In a sense MOPED is much more powerful than it needs to be,
but this is hardly a criticism.  With MOPED and the advances of
\scite{TZH01}, parameter estimation is accelerated by a useful factor
of $\sim 10^3$, and we can be fairly certain that the data
processing element will be dominated by other steps in the analysis
pipeline.

The speed of MOPED may influence the analysis strategy; if the
likelihood evaluation is slow in comparison with theoretical power
spectrum generation, then one can compute the power spectra
`on-the-fly' in a search for the maximum likelihood.  Given that the
position is now reversed, there is a case for creating grids of
theoretical models in the years before launch of Planck.  If storage
space becomes a limiting issue, one can store the expected MOPED
components for each model, rather than the full $C_\ell$, with a
compression factor $>100$.  However, there may well still be a case
for less rigid searches of parameter space, such as Markov Chain Monte
Carlo methods \cite{Christensen2001}, since they can simultaneously
estimate the shape of likelihood surface around the peak, as well as
finding the peak itself.  MOPED can be combined with such methods to
advantage.  Finally we note that, for current experiments, data
compression is not necessary, as there are relatively few band-power
estimates available.

\noindent{\bf Acknowledgments}

\noindent Computations were made partly using Starlink facilities.

\bibliographystyle{mnras}
\bibliography{general}

\end{document}